\documentclass[twocolumn]{aastex63}

\usepackage{multirow}

\newcommand{\QSO}     {COOL\,J0542$-$2125}
\newcommand{\QSOLone}    {SDSS\,J1004$+$4112}
\newcommand{\QSOLtwo}    {SDSS\,J1029$+$2623}
\newcommand{\QSOLthree}     {SDSS\,J2222$+$2745}
\newcommand{\QSOLfour}     {SDSS\,J0909$+$4449}
\newcommand{\QSOLfive}     {SDSS\,J1326$+$4806}

\newcommand{\Lenstool}{{\tt{Lenstool}}}

\received{ }
\revised{ }
\accepted{ }
\submitjournal{ApJ}

\shorttitle{A 25\farcs 9 Wide-Separation Lensed Quasar}
\shortauthors{Martinez et al.}
\graphicspath{{./}{figures/}}

\begin{document}

\title{COOL-LAMPS III: Discovery of a 25\farcs9 Separation Quasar Lensed by a Merging Galaxy Cluster \footnote{This paper includes data gathered with the 6.5m Magellan Telescopes located at Las Campanas Observatory, Chile, and the 2.5m Nordic Optical Telescope located at Rocque de los Muchachos Observatory, La Palma.}}

\correspondingauthor{Michael N. Martinez}
\email{mnmartinez@wisc.edu}
\author[0000-0002-8397-8412]{Michael N. Martinez}
\altaffiliation{Current affiliation: Department of Physics, University of Wisconsin, Madison, 1150 University Avenue, Madison, WI 53706, USA}
\affiliation{Department of Astronomy and Astrophysics, University of
Chicago, 5640 South Ellis Avenue, Chicago, IL 60637, USA}
\author[0000-0003-4470-1696]{Kate A. Napier}
\affiliation{Department of Astronomy, University of Michigan, 1085 S. University Ave, Ann Arbor, MI 48109, USA}
\author[0000-0001-9978-2601]{Aidan P. Cloonan}
\affiliation{Department of Astronomy and Astrophysics, University of
Chicago, 5640 South Ellis Avenue, Chicago, IL 60637, USA}
\author[0000-0002-1106-4881]{Ezra Sukay}
\affiliation{Department of Astronomy and Astrophysics, University of
Chicago, 5640 South Ellis Avenue, Chicago, IL 60637, USA}
\author[0000-0003-2294-4187]{Katya Gozman}
\affiliation{Department of Astronomy and Astrophysics, University of
Chicago, 5640 South Ellis Avenue, Chicago, IL 60637, USA}
\author[0000-0001-5931-5056]{Kaiya Merz}
\affiliation{Department of Astronomy and Astrophysics, University of
Chicago, 5640 South Ellis Avenue, Chicago, IL 60637, USA}
\author[0000-0002-3475-7648]{Gourav Khullar}
\affiliation{Department of Astronomy and Astrophysics, University of
Chicago, 5640 South Ellis Avenue, Chicago, IL 60637, USA}
\affiliation{Kavli Institute for Cosmological Physics, University of
Chicago, 5640 South Ellis Avenue, Chicago, IL 60637, USA}
\affiliation{Department of Physics and Astronomy and PITT PACC, University of Pittsburgh, Pittsburgh, PA 15260, USA}
\author[0000-0003-1266-3445]{Jason J. Lin}
\affiliation{Department of Astronomy and Astrophysics, University of
Chicago, 5640 South Ellis Avenue, Chicago, IL 60637, USA}
\author[0000-0001-9225-972X]{Owen S. Matthews Acu\~{n}a}
\affiliation{Department of Astronomy and Astrophysics, University of
Chicago, 5640 South Ellis Avenue, Chicago, IL 60637, USA}
\author{Elisabeth Medina}
\affiliation{Department of Astronomy and Astrophysics, University of
Chicago, 5640 South Ellis Avenue, Chicago, IL 60637, USA}
\author[0000-0002-9142-6378]{Jorge A. Sanchez}
\affiliation{Department of Astronomy and Astrophysics, University of
Chicago, 5640 South Ellis Avenue, Chicago, IL 60637, USA}
\author[0000-0002-2358-928X]{Emily E. Sisco}
\affiliation{Department of Astronomy and Astrophysics, University of
Chicago, 5640 South Ellis Avenue, Chicago, IL 60637, USA}
\author[0000-0001-8008-7270]{Daniel J. Kavin Stein}
\affiliation{Department of Astronomy and Astrophysics, University of
Chicago, 5640 South Ellis Avenue, Chicago, IL 60637, USA}
\author[0000-0001-6584-6144]{Kiyan Tavangar}
\affiliation{Department of Astronomy and Astrophysics, University of
Chicago, 5640 South Ellis Avenue, Chicago, IL 60637, USA}
\author[0000-0002-7868-9827]{Juan Remolina Gonz\`{a}lez}
\affiliation{Department of Astronomy, University of Michigan, 1085 S. University Ave, Ann Arbor, MI 48109, USA}
\author[0000-0003-3266-2001]{Guillaume Mahler}
\affiliation{Department of Astronomy, University of Michigan, 1085 S. University Ave, Ann Arbor, MI 48109, USA}
\affiliation{Department of Physics, Durham University, South Road, Durham DH1 3LE}
\author[0000-0002-7559-0864]{Keren Sharon}
\affiliation{Department of Astronomy, University of Michigan, 1085 S. University Ave, Ann Arbor, MI 48109, USA}
\author[0000-0003-2200-5606]{H{\aa}kon Dahle}
\affiliation{Institute of Theoretical Astrophysics, University of Oslo, P.O. Box 1029, Blindern, NO-0315 Oslo, Norway}
\author[0000-0003-1370-5010]{Michael D. Gladders}
\affiliation{Department of Astronomy and Astrophysics, University of Chicago, 5640 South Ellis Avenue, Chicago, IL 60637, USA}
\affiliation{Kavli Institute for Cosmological Physics, University of Chicago, 5640 South Ellis Avenue, Chicago, IL 60637, USA}







\begin{abstract}
In the third paper from the COOL-LAMPS Collaboration, we report the discovery of COOL J0542-2125, a gravitationally lensed quasar at $z=1.84$, observed as three images due to an intervening massive galaxy cluster at $z=0.61$.  The lensed quasar images were identified in a search for lens systems in recent public optical imaging data and have separations on the sky up to 25\farcs 9, wider than any previously known lensed quasar. The galaxy cluster acting as a strong lens appears to be in the process of merging, with two sub-clusters separated by $\sim 1$ Mpc in the plane of the sky, and their central galaxies showing a radial velocity difference of $\sim 1000$\,km/s. Both cluster cores show strongly lensed images of an assortment of background sources, as does the region between them.  A preliminary strong lens model implies masses of $M(<250\ \rm{kpc}) = 1.79^{+0.16} _{-0.01} \times 10^{14} M_{\odot}$ and  $M(<250\ \rm{kpc}) = 1.48^{+0.04}_{-0.10} \times 10^{14} M_{\odot}$ for the East and West sub-clusters, respectively.
This line of sight is also coincident with a ROSAT ALL-sky Survey source, centered between the two confirmed cluster halos reminiscent of other major cluster-scale mergers. 
Archival and new follow-up imaging show flux variability in the quasar images of up to 0.4 magnitudes within $\sim 1$ year, and new multi-color imaging data reveal a $2\sigma$ detection of the underlying quasar host.
A lens system with this configuration offers rare opportunities for a range of future studies of both the lensed quasar and its host and the foreground cluster merger causing the lensing.


\end{abstract}



\section{Introduction} \label{sec:intro}
Since the initial discovery of a gravitationally lensed quasar by \cite{walsh79}, more than 300 quasars have been found to be strongly lensed by an intervening mass along the line of sight \citep{lemon19, lemon22}.
In the vast majority of these cases, the object acting as a lens is an individual galaxy, producing typical image separations of 1-2\arcsec. Only five cases with image separations larger than 10\arcsec, corresponding to a cluster-scale lens mass M$_{200} \gtrsim 10^{14}\,M_{\sun} $, have so far been reported in the literature \citep{inada03,inada06,J2222,shu18,shu19}. These five cases have maximum image separations in the range 14\arcsec - 22\arcsec. Numerical simulations  \citep{hilbert08,robertson20} predict a significantly higher abundance of cluster-mass lenses than seen in the current sample of lensed quasars. \cite{robertson20} found that for a typical quasar source redshift of $z_s=2$, gravitational lens masses of M$_{200} > 10^{14}\,M_{\sun} $ will contribute 25\% of the total lensing cross section. This implies that the current sample is biased towards galaxy-scale lens masses and that the true fraction of quasars lensed by cluster-scale masses could be an order of magnitude larger than suggested by the currently known lensed quasars. 

\begin{figure*}
    \centering
    \includegraphics[width=500px]{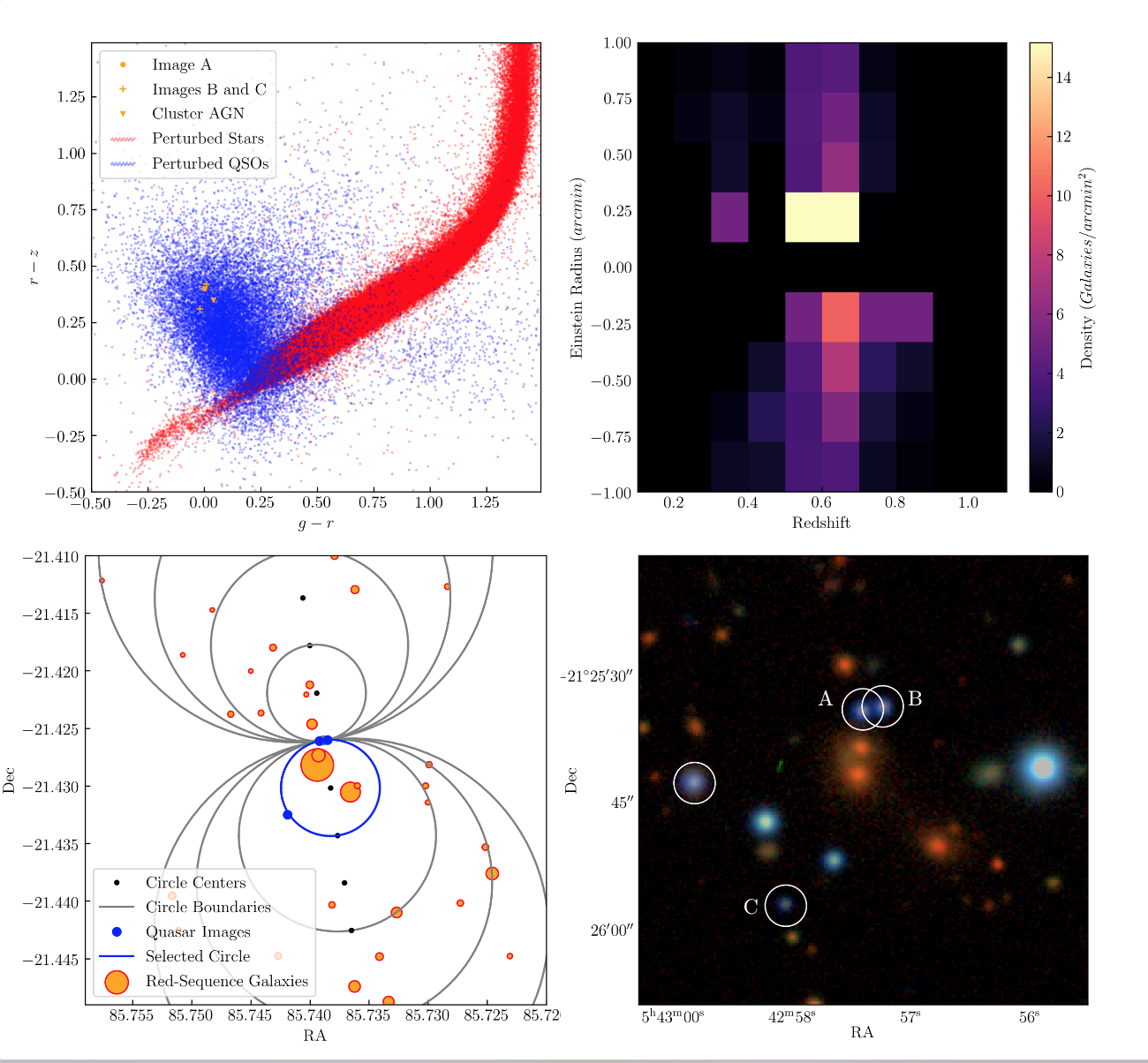}
    \caption{Selected images from the search process. Top Left: A perturbed color-color diagram of known stars and known quasars, from the SDSS dataset. Quasar images are shown, as well as the false-positive in-cluster AGN seen in the bottom right panel. Top Right: Map of early-type galaxy (ETG) number density in a range of redshift slices and prospective Einstein radii. Bottom Left: On-sky map showing the ETG's and circles representing Einstein radii from the previous panel. Bottom Right: DECaLS image of the area around COOL J0542-2125, with objects identified as possible lensed quasar images circled. Note the cluster-redshift AGN contaminant at the left of the image (unlabeled circle).}
    \label{fig:search}
\end{figure*}

The quasars lensed by clusters constitute a distinct population of lens systems that make them particularly advantageous for studies of both the source quasar and its host galaxy, as well as the lensing cluster and potentially cosmology. The large-separation images probe the source from multiple viewing angles that differ enough to reveal the 3D structure of quasar outflows \citep{misawa13,misawa14,misawa16}. The high magnification and wide image separation from the foreground lens optimizes conditions for resolved studies of the quasar host galaxy, as the host can be disentangled from the bright quasar by the lensing effect \citep{bayliss17}. Wide-separation lensed quasars can also be excellent targets for precise measurements of the quasar's supermassive black hole mass through reverberation mapping, as demonstrated by \cite{williams21A,williams21B}.
Measurements of the time delays between the different quasar images constrains a different derivative of the lensing potential than the image positions or the combination of positions and magnification provide. Together with redshifts and positions of other multiply lensed background sources, this constrains the mass distribution of the cluster \citep{sharon17,oguri13}. 

Measurements of time delays between the different quasar images lensed by a cluster \citep{Fohlmeister07,Fohlmeister08, fohlmeister13, Dahle15, munoz22} can in principle also provide constraints on the Hubble parameter $H_0$ using the method of \cite{refsdal64}. Through its dependence on the lensing mass distribution in the dark matter dominated regime, the use of clusters for such studies largely avoids issues with systematic uncertanties produced by the baryonic mass component which are difficult to resolve observationally for galaxy-scale lenses \citep{2021MNRAS.501.5021K,2022A&A...663A.179V}. Nevertheless, the extant literature on $H_0$ measurements from strong lensing time-delays on cluster scales consists of one measurement of a lensed quasar \citep{kumar15} and one cluster-scale lensed supernova \citep{refsdalH0}. A larger sample of quasars lensed by clusters could be used to constrain $H_0$ through time-delays, as well as models for cosmology and structure formation through their overall abundance and the distribution of different lensing multiplicities and image configurations.  


In this paper we present a new discovery from a catalog-based search of the Dark Energy Camera Legacy Survey (DECaLS) \citep{legacysurveysoverview}. The search algorithm is described in Section \ref{sec:discovery}. We describe follow-up spectroscopy that confirms this candidate as a gravitationally lensed quasar in Section \ref{subsec:observations}. In Section \ref{sec:analysis} we show that the lensing cluster is one of two sub-clusters comprising a merging system. 
Using photometry based on archival and new imaging data, we detect significant intrinsic variability in the lensed source, leading to an initial constraint on the time delay between the two brightest quasar images. We use these ground-based data to model the mass distribution of both sub-clusters and detect the quasar host-galaxy in the two brighter lensed images. We summarize and report on further opportunities for study in Section \ref{sec:discussion}.  
For all geometry and cosmology calculations we assume a flat $\Lambda$CDM cosmology with $\Omega_{\Lambda}$ = 0.7, $\Omega_{m}$ = 0.3, and \textit{H}$_{0}$ = 70 km s$^{-1}$ Mpc$^{-1}$. Magnitudes are reported in the AB system.

\begin{figure*}
    \centering
    \includegraphics[width=500px]{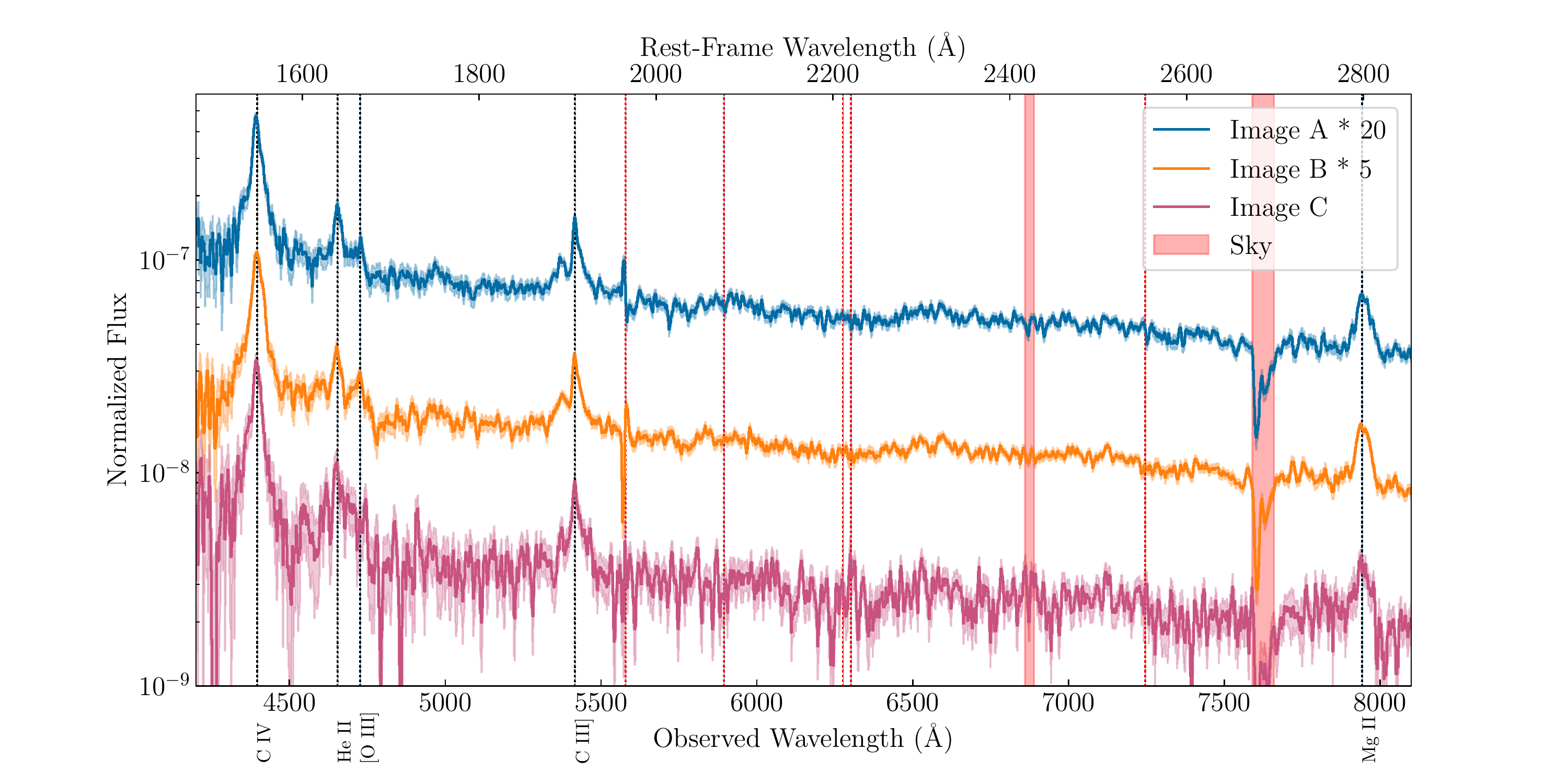}
    \caption{LDSS3 spectra of the three images of \QSO. Data were smoothed with a moving average filter and normalized according to r-band magnitude. The spectra of Images A and B were shifted by a constant (see legend) to improve readability, and all wavelengths have been corrected to vacuum.}
    \label{fig:spec}
\end{figure*}

\section{Discovery} \label{sec:discovery}

\QSO\ was discovered as part of the ChicagO Optically-selected strong Lenses - Located At the Margins of Public Surveys (COOL-LAMPS) collaboration, in a search for wide-separation lensed quasars utilizing archival data from DECaLS Data Release 8 (DR8).

Unlike other lens-finding efforts by this collaboration \citep{CJ1241} the search for wide-separation lensed quasars was primarily catalog-based, with imaging data only examined at a late stage in the process.
We began by limiting our search area to regions near luminous red galaxies (LRGs) which we selected using galaxy color-magnitude cuts. LRGs are often found in clusters and are markers of large dark matter halos \citep{2000AJ....120.2148G} and so are an obvious choice when searching for cluster-scale lenses.
To save computational time when analyzing the multi-million-object DR8 catalog, the redshift of these galaxies and their orientation relative to candidate quasars was not taken into account at this stage.
We assumed any lensed quasar would be cataloged in DECaLS as a point source, and therefore selected any well-measured point source in the aforementioned fields as a candidate.
To select the subset of point sources most likely to be quasars, we adopted a probabilistic approach applied in color-color space. For each candidate point source, we considered its colors against two samples: a set of spectroscopically confirmed quasars from the Sloan Digital Sky Survey (SDSS), and bright but unsaturated DECaLS point sources, the vast majority of which are stars.
We perturbed the magnitudes of our quasar and star test points by the uncertainty of the candidate DECaLS source, and then counted the number of stars and quasars that fell within a small circle around the candidate point source in $g$-$r$ and $r$-$z$ color-color space.
The proportion of quasar to stellar test points gives a relative likelihood of that source being a quasar rather than a star. This probability is not absolute, since no absolute density of stars and quasars is used.   
An example color-color diagram with the perturbed points is shown in the top left panel of Figure \ref{fig:search}.
After the initial photometric analysis, we matched all high-likelihood point pairs on the sky, with a maximum image separation for a match set at 30", which was 7.5" wider than the published widest known quasar lens, \QSOLtwo\,\citep{inada03}.

We further analyzed the photometry of these pairs to increase the likelihood that a given on-sky pair of quasar candidates was indeed two images of the same object. Quasars vary significantly in color and brightness over time \citep{quasarcolors},
and time delays inherent in gravitational lensing imply real differences in quasar image colors at a given observation epoch.
Thus, a strict match of colors or magnitudes within a possible quasar pair would be unsuitable for a search.
Taking this into consideration, we examined the known sample of cluster-scale lensed quasars and used their color differences in DECaLS to create a conservative cut to object pairs with drastically different colors. We considered a candidate quasar pair more likely to be lensed if its components were within 1 magnitude of each other in $grz$-space.
Altogether, this catalog-focused analysis produced hundreds of thousands of candidates, which, while an improvement from the half-billion point sources in DECaLS, was still far too many to inspect visually.

To further refine our search for gravitationally lensed quasars, we utilized a geometric approach modeling each image in a candidate pair as lying on a common Einstein Ring. While there are an infinite number of circles that share these two points, we selected only ten - five on each side of the candidate quasar pair - with radii ranging from half of the distance between pair members to 1 arcminute. An example of this circle selection process is illustrated in the bottom left panel of Figure \ref{fig:search}.

Red-sequence galaxies are excellent markers of cluster- and group-scale dark matter halos \citep{2000AJ....120.2148G}.
While the first stage of the search used an LRG catalog to select search areas via a simple sky coordinate match, the reduced size of the candidate quasar pair dataset allowed for a more detailed analysis in tandem with the Einstein Ring approach.
Using DECaLS DR8 catalog data, we calculated number densities of red-sequence galaxies at a range of redshifts within each chosen circle. This produces an overdensity ‘map' in redshift and radius, an example of which can be seen in the top right panel of Figure \ref{fig:search}. A spatial example showing the circles (but not accounting for redshift) is displayed in the bottom left panel.
We compared these density maps to those produced by a background of randomly spaced pairs of points on the sky, and selected pairs with higher density at a given radius and redshift than that background, resulting in a few thousand final lensed quasar candidates.

These were visually inspected by two co-authors and given a score on a scale of 0 to 3, with 3 being a definite lensed quasar in the scorer's opinion. High scoring candidates were gathered and examined further, utilizing other archival surveys such as unWISE \citep{2014AJ....147..108L} and GALEX \citep{galex}, to create a high-priority candidate list for spectroscopic follow-up, of which DECALS J0542-2125 was one.

\begin{figure*}
    \centering
    \includegraphics[width=500px]{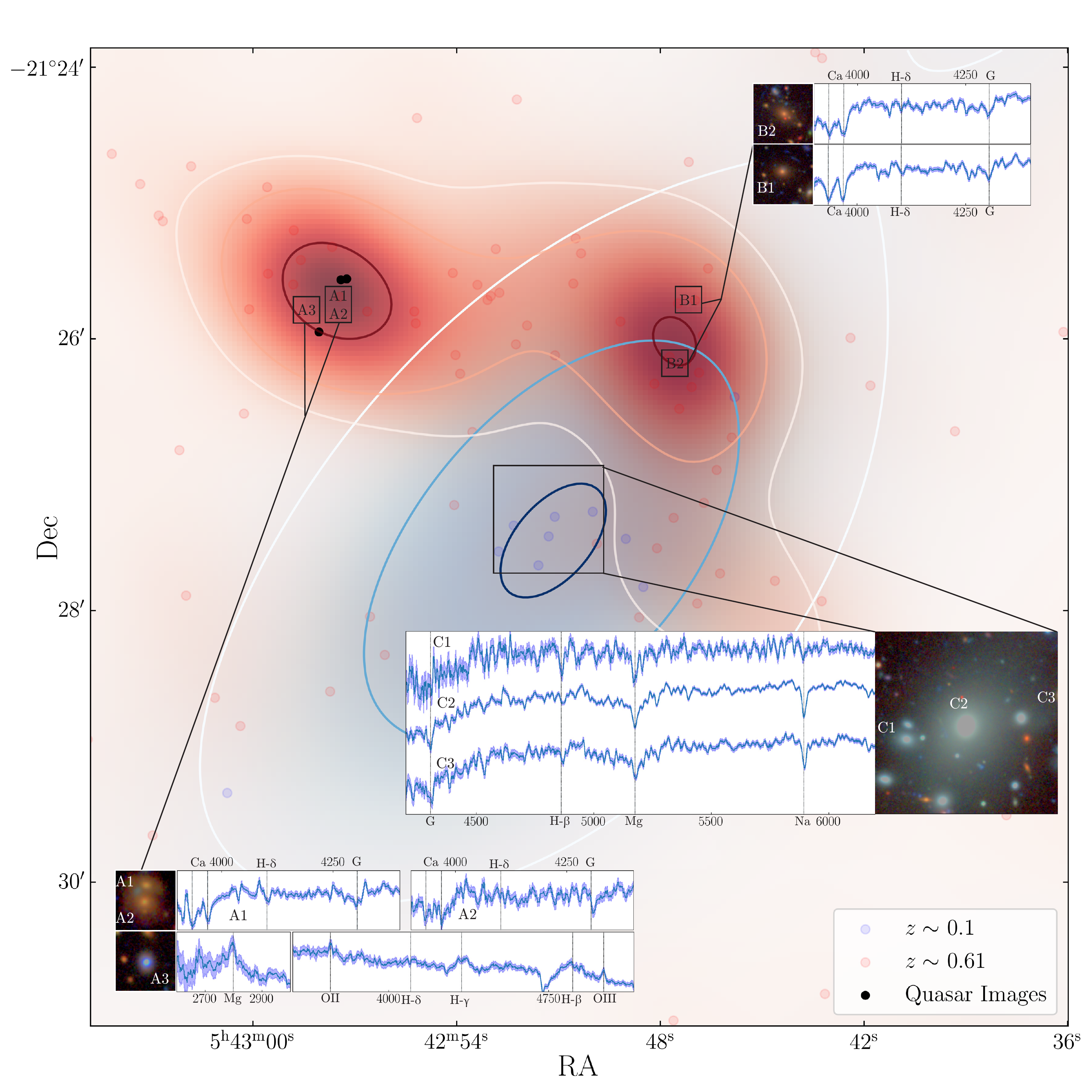}
    \caption{Background: Map of the surrounding galactic structure around COOL J0542$-$2125 at the lens redshift and in the foreground. Contour lines correspond to the equivalent of Gaussian 1$\sigma$, 2$\sigma$, and 3$\sigma$ significance levels. Insets: Multiband imaging and spectra of objects observed with the LDSS3 spectrograph. Spectra are rest-frame and smoothed with a moving-average filter, with recognized lines marked. Color $grz$ images were taken with LDSS3 and were created using the \cite{Lupton_2004} intensity rescaling method.}
    \label{fig:density}
\end{figure*}

\begin{table}
\centering
\begin{tabular}{c c c c}
 \hline
 ID & R.A. (deg) & Decl (deg) & \textit{z}\\ [0.5ex] 
 \hline
Image A & 85.7392 & -21.4261 & $1.83833 \pm 0.00086$\\
Image B & 85.7385 & -21.4260 & $1.83833 \pm 0.00086$\\
Image C & 85.7419 & -21.4325 & $1.83833 \pm 0.00086$\\
A1 & 85.7393 &  -21.4274 & $0.6093 \pm 0.0005$ \\ 
A2 & 85.7394 &  -21.4282 & $0.6133 \pm 0.001$ \\
A3 & 85.7451  & -21.4285 & $0.611 \pm 0.002$ \\
B1 & 85.6957 & -21.4286 & $0.6137 \pm 0.0007$ \\
B2 & 85.6969 & -21.4264 & $0.6158 \pm 0.0006$ \\
C1 & 85.7180 & -21.4563 & $0.10044 \pm 0.0005$ \\
C2 & 85.7130 & -21.4552 & $0.09681 \pm 0.00015$ \\
C3 & 85.7083 & -21.4546 & $0.09641 \pm 0.00012$ \\
\end{tabular}
\caption{\label{tab:density_redshifts} Coordinates and redshifts of the sources targeted for spectroscopy with LDSS3 and shown in Figures \ref{fig:spec} and \ref{fig:density}.}
\end{table}

\subsection{Follow-Up Observations} \label{subsec:observations}
Two images of \QSO, labeled as A \& B in Figure \ref{fig:search}, were observed spectroscopically on October 21, 2020 using the LDSS3C spectrograph on the Magellan 2 Clay telescope.
The total integration time was 600 seconds, in photometric conditions, using a 1\farcs25 wide longslit, dispersed by the VPH-ALL grism; this yields a spectral resolution of $R\sim500$ for an unresolved source filling the slit. However, the seeing at the time of observation was 0\farcs55, and so the effective resolution of the resulting spectrum is somewhat finer, i.e. $R\sim1200$. These initial spectra confirmed images A\&B as two images of the same background quasar.  A further spectroscopic observation, totaling 1200 seconds and using the same setup on October 22, 2020, targeted the suspected third image (image C in Figure \ref{fig:search}) under similar conditions.
This latter slit placement also targeted image A, and contains sufficient flux to deduce a redshift for the two bright cluster galaxies between images A and C.  One final spectroscopic observation totaling 240 seconds was taken in twilight on the same night, targeting the three other bright blue point sources - including one tagged by our analysis as a likely AGN (See Figure \ref{fig:search}) -  near the lensing cluster center. Imaging of \QSO~was also acquired on this second night of observing, totaling 450 seconds in the $z$ filter, and 540 seconds in each of the $g$ and $r$ filters, in $\sim$0\farcs5 seeing conditions.

Initial analysis of the DECaLS images and photometry indicated a complex mass field (see \S \ref{subsec:density} below) and motivated additional spectroscopy and imaging. Three further longslit positions were observed on the nights of November 13 and 14, 2020. 
Two of these were placed to measure other mass structures near the lensed quasar images, and the third was a re-observation of images A\&B using an order separating filter to cross check the identification of several faint features in the initial spectrum as second-order lines from the extreme blue. The observing setup was otherwise the same as the initial observations described above. \QSO~ was also imaged for a total of 720 second in each of the $grz$ filters on November 13, 2020, again in $\sim$0\farcs5 conditions.

\section{Analysis} \label{sec:analysis}

\subsection{Galaxy Clusters in the COOL J0542–2125 Field\label{subsec:density}}
The DECaLS $grz$ color image \citep{legacysurveysoverview} of the area surrounding \QSO\ shows two obvious large-scale mass structures near the lens, highlighted by early-type galaxies with colors consistent with expectations for red-sequence galaxies.
The first is a foreground cluster with redshift $z \sim 0.1$, while the latter is the \QSO\ lensing cluster itself and a neighboring cluster at the same redshift.
Figure \ref{fig:density} displays a map of these overdensities, estimated with a Gaussian kernel weighted by galaxy flux, and with a kernel FWHM of 650 kpc at the two redshifts shown. The two peaks of the $z = 0.61$ density map are separated on the sky by an angular distance of 2.37 arcmin (0.948 Mpc), and the two BCGs are separated by 2.42 arcmin (0.968 Mpc).

\subsection{Single-Epoch Images}\label{subsec:epoch}
\begin{figure}
    \centering
    \includegraphics[width=241px]{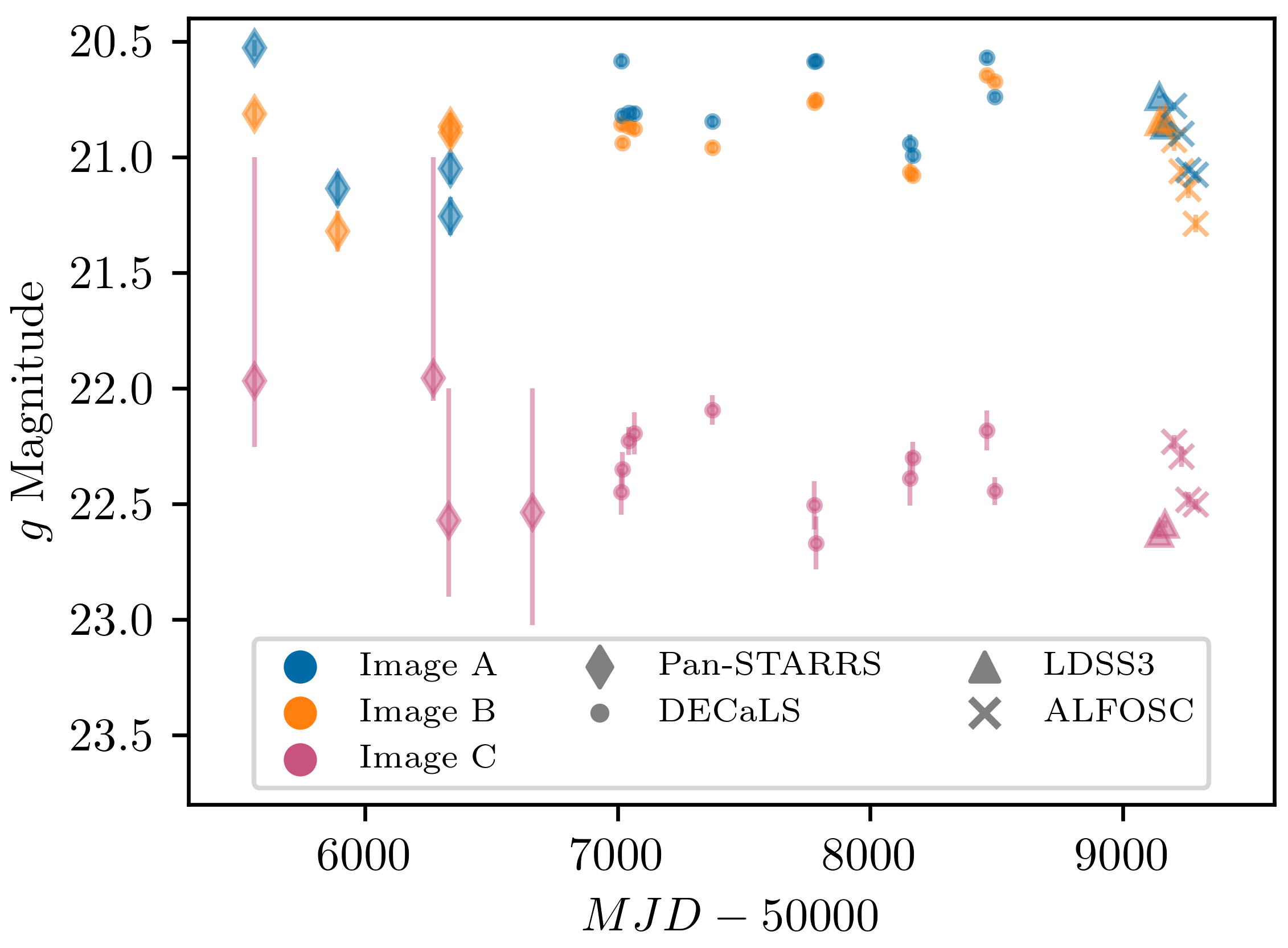}
    \caption{Single-epoch photometry of the three images of COOL J0542$-$2125 in the $g$ band. Blue, orange, and pink represent images A, B, and C respectively. Diamonds represent archival PAN-STARRS photometry, and circles DECaLS DR9. Triangles represent Magellan 2 Clay LDSS3 photometry, and x's represent Nordic Optical Telescope ALFOSC photometry. The data indicate that the quasar shows high variability, and is suitable for time delay and reverberation studies.}
    \label{fig:epoch}
\end{figure}

Using archival single-epoch imaging from the DECaLS and PAN-STARRS surveys, as well as that from the October 2020 observations, we constructed light curves for the three confirmed images of \QSO, as can be seen in Figure \ref{fig:epoch}. This figure also includes 4 epochs of follow-up photometry using the Alhambra Faint Object Spectrograph and Camera (ALFOSC) at the 2.56 m Nordic Optical Telescope. This limited light curve is insufficient to measure a time delay for the lensed system, but does establish the quasar's high variability, indicating such a study is achievable with more observations. Our limited time-series data does show a tight correlation between images A and B, suggesting a short time delay between those two images - as would be expected for their small ($\sim 2.4"$) separation. The recent measurement of a 6-year time delay for a similarly wide-separation quasar lens, \QSOLone\
\citep{munoz22}, suggests the delay between the close images and image C could be on the order of several years.

\subsection{Lens Modeling}\label{subsec:lensmodel}
We construct a preliminary strong lensing model for the foreground structure lens of \QSO\ from the existing ground-based imaging and spectroscopy. 
Strong lens modeling requires observational constraints in the form of sets of multiple images of the same background sources. The three images of the lensed quasar, which we spectroscopically confirmed as multiple images of the same source (section~\ref{subsec:observations}), are used as lensing constraints. 
We identify several other lensed background sources in the field as described below. We estimate their photometric redshifts using the Markov Chain Monte Carlo (MCMC)-based stellar population synthesis (SPS) and parameter inference code, \texttt{Prospector} \citep{prospector_sps}.
\texttt{Prospector} is based on the \texttt{Python-FSPS} framework, with the MILES stellar spectral library and the MIST set of isochrones \citep{2010ApJ...712..833C,2017ApJ...837..170L,2013PASP..125..306F,2011A&A...532A..95F,Choi_2016}. We use a single burst star formation history model (corresponding to a unique formation episode of simple stellar populations in this galaxy at burst age t$_{age}$). This model also used as free parameters the redshift, total stellar mass formed ($M_{\star}$) and  metallicity log$(Z/Z_{\odot})$. With suitable priors corresponding to a quiescent galaxy, we sample posterior distributions using MCMC.

\begin{figure*}
    \centering
    \includegraphics[width=500px]{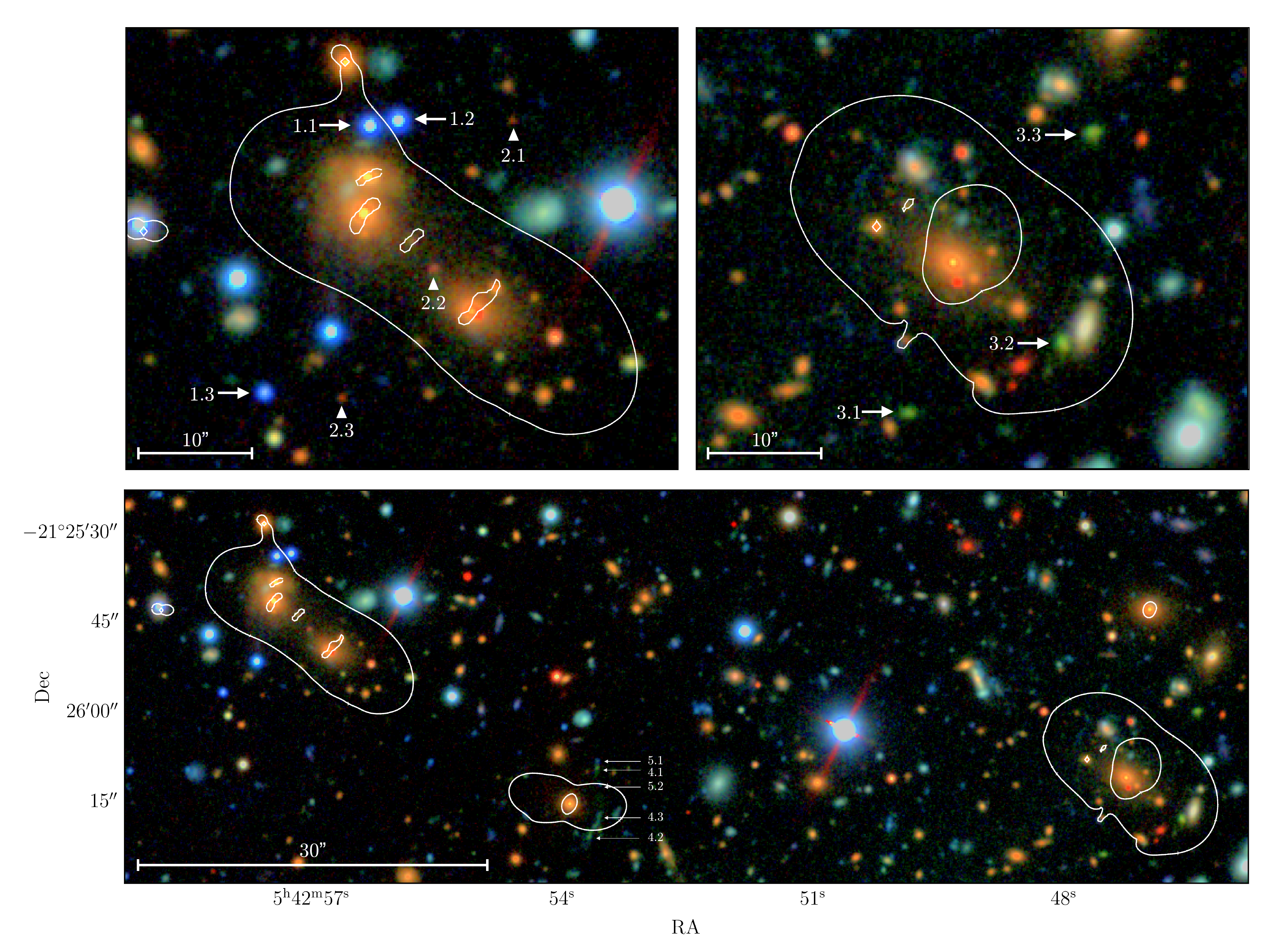}
    \caption{COOL J0542$-$2125 and its surroundings, with zoomed-in views of the main subclusters in the top panels. Lensing critical curves are shown in white. The quasar lensing ``East" cluster is visible on the left, with the three quasar images marked by arrows. Another object used to constrain the model, at a photometric redshift of $z\sim 1.7$, is marked with triangles. The critical curves shown correspond to the quasar redshift at $z=1.84$. On the right, the ``West" cluster is shown. Critical curves are drawn at $z=4.3$, corresponding to the green $z\sim 4.3$ object used to constrain the model and marked with arrows. In the bottom center of the wide-field view, a third lens system of unknown source redshift is visible between the two main subclusters, with critical curves drawn at the quasar redshift of 1.84. Components of the arc used to constrain the model are labeled. Images are $grz$ and were taken with LDSS3 and rendered using the \cite{Lupton_2004} intensity rescaling method.}
    
    \label{fig:lensmodel}
\end{figure*}

We identify a set of three images with similar color in the East sub-cluster, with an estimated photometric redshift of $z = 1.7 \pm 0.1$ (triangles pointing to three red sources in the top-left panel of Figure~\ref{fig:lensmodel}).
In the West sub-cluster, we identify a set of three images of a background galaxy with distinctive green color in the $grz$ image (top-right panel of Figure~\ref{fig:lensmodel}). Its estimated photometric redshift is $z = 4.3 \pm 0.1$.
In the space between the two sub-clusters, we identify two families of multiple images.  We use the aforementioned images as constraints. Their positions and spectroscopic or photometric redshifts are listed in Table~\ref{tab:lens_mod_const}.

We compute the strong lensing model using the public software \Lenstool\ \citep{2007NJPh....9..447J}.  \Lenstool\ uses a parametric modeling algorithm which describes the mass distribution as a combination of halos, each defined by a set of parameters.  \Lenstool\ utilizes MCMC sampling of the parameter space and identifies the best-fit model as the one that minimizes the scatter between the observed and model-predicted image locations in the image plane.  


We model the lens plane with three mass halos representing the dark matter mass distribution of the sub-clusters and correlated large scale structure, supplemented with mass halos assigned to individual cluster-member galaxies.  Both the cluster-scale and galaxy-scale potentials are modeled as a pseudo-isothermal ellipsoidal mass distribution (PIEMD, also known as dPIE; \cite{2007NJPh....9..447J,2007arXiv0710.5636E}) described by seven parameters: position \textit{x}, \textit{y}; ellipticity, \textit{e} = (a$^{2}$-b$^{2}$)/(a$^{2}$+b$^{2}$), where \textit{a} and \textit{b} are the semi-major and semi-minor axes, respectively; position angle \textit{$\theta$}, measured north of west; core radius r$_{core}$; cut radius r$_{cut}$; and effective velocity dispersion $\sigma_{0}$.  For the cluster-scale halos, we allow all of the parameters to vary, except for r$_{cut}$, which is fixed at 1500 kpc.  This is because, for a typical cluster, the cut radius extends further than the radius at which lensing evidence is found, prohibiting it from being constrained by the model.  
The galaxy scale halos' normalization and radii parameters are scaled to their observed luminosity (a description of the scaling relations can be found in \cite{10.1111/j.1365-2966.2004.08449.x}). The positional parameters ($x$, $y$, $e$, $\theta$) are measured with \texttt{SourceExtractor} \citep{sextractor} and fixed.   

To select cluster member galaxies, we generated photometric catalogs of all non-stellar objects in the field in \textit{grz} from LDSS3 using \texttt{SourceExtractor} with a detection threshold of 5$\sigma$ and a deblend parameter of 0.001 in dual-image mode. Galaxies were identified in the \textit{z}-band and colors measured in the \textit{r}- and \textit{g}- bands within the same aperture. Cluster members were selected by their color relative to the red sequence in a color-magnitude diagram (\textit{r-z} vs. \textit{z}) \citep{2000AJ....120.2148G}.  The \textit{z}-band mag\_auto magnitude and the shape parameters (\textit{a}, \textit{b}, \textit{$\theta$}) were used in our galaxy catalog as the \textit{z}-band best samples the stellar mass at the cluster redshift. 

The critical curves of the best-fit lens model are overplotted on the $grz$ LDSS3 image of the field in Figure~\ref{fig:lensmodel}. To guide the eye to the relevant lensing configuration in each sub-cluster, we plot the critical curve for a source at the quasar redshift in the East half of the field, and for a source at $z=4.3$ in the West. The preliminary lens model is somewhat under-constrained considering the complexity of the lens plane. Significant improvement can be made once space-based imaging is obtained, which will enable multiplexing the constraints in the current lensed images by using the currently unresolved substructure; confirming candidate lensed feature to be used as constraints; and identifying additional lensed galaxies. 

We use the best fit lens model to measure the projected mass density enclosed within 250 kpc from the BCG of each of the sub-clusters, and find $M(<250\,kpc) = 1.79^{+0.16}_{-0.01} \times 10^{14} M_{\odot}$ and  $M(<250\,kpc) = 1.48^{+0.04}_{-0.10} \times 10^{14} M_{\odot}$ for the East and West sub-clusters, respectively. Uncertainties are derived from a suite of 100 random models from the MCMC sampling of the parameter space, and the errors represent $1-\sigma$.

\begin{table}
\begin{tabular}{c c c c} 
 \hline
 ID & R.A. (J2000) & Decl. (J2000) & \textit{z}\\ [0.5ex] 
 \hline
1.1 & 85.73924927 &  -21.42612745 & 1.84 \\
1.2 & 85.73851675 &  -21.42600703 & 1.84 \\ 
1.3 & 85.74191404  & -21.43245443 & 1.84 \\
2.1 & 85.7356128 & -21.4260133 & $1.7 \pm 0.1$\\
2.2 & 85.7376021 & -21.4295212 & $1.7 \pm 0.1$ \\
2.3 & 85.7399502 & -21.4325734 & $1.7 \pm 0.1$ \\
3.1 & 85.6980251 & -21.4399683 & $4.3 \pm 0.1$ \\
3.2 & 85.6940618 & -21.4383539 & $4.3 \pm 0.1$ \\
3.3 & 85.6933073 & -21.4333820 & $4.3 \pm 0.1$ \\
4.1 & 85.7232636 & -21.4360147 & \nodata \\ 
4.2 & 85.7236361 & -21.4391366 & \nodata \\
4.3 & 85.7231149 & -21.4381354 & \nodata \\
5.1 & 85.7232337 & -21.4356442 & \nodata \\
5.2 & 85.7231199 & -21.4367619 & \nodata \\ 
\end{tabular}
\caption{\label{tab:lens_mod_const} Images used as constraints in lens model. All redshifts shown are photo-z's besides objects 1.x, the spectroscopically confirmed quasar images.}
\end{table}


\begin{table*}
\centering
\begin{tabular}{c c c c c c c} 
 \hline
 \multirow{2}{2em}{\centering ID} & \multirow{2}{4em}{\centering Lens Redshift} &
 \multirow{2}{5em}{\centering Source Redshift} & \multirow{2}{5em}{\centering Number of Images} & \multirow{2}{7em}{\centering Max Image Separation (")} & \multirow{2}{7em}{\centering Brightest Image (DECam $g$)} & \multirow{2}{7em}{\centering Discovery Reference}\\ [0.5ex] 
 \\
 
 \hline
\QSO & 0.61 & 1.84 & 3 & 25.9 & 20.7 & This work \\ 
\QSOLtwo & 0.584 &  2.199 & 3 & 22.54 & 18.5 & \cite{inada06} \\
\QSOLfive & 0.396 & 2.078 & 2 & 21.1 & 20.2 & \cite{shu19} \\
\QSOLthree & 0.49 & 2.82 & 6 & 15.1 & 20.4 & \cite{J2222} \\
\QSOLone & 0.68 & 1.74 & 5 & 14.72 & 19.7 & \cite{inada03} \\
\QSOLfour & 0.9 & 2.788 & 3 & 14.0 & 21.5 & \cite{shu18}\\

\end{tabular}
\caption{\label{tab:lensed_qsos} Currently published cluster-scale lensed quasars, ordered by maximum image separation (shown in arcseconds).}
\end{table*}

\section{Discussion and Future Work} \label{sec:discussion}
The COOL J0542$-$2125 lensed quasar system, as well as the greater galactic structure around it, present a remarkable opportunity for further study. 

\subsection{Time-Domain Observations}
\label{qsodiscussion}
As mentioned in Section \ref{subsec:epoch}, \QSO\,displays significant variability in brightness across the three confirmed quasar images. An exact time delay for each lensed image is beyond the scope of this paper and the data utilized in it, but from the image separation this delay could be multiple years, requiring extensive follow-up observations that are already underway. Additionally, COOL J0542$-$2125 is the only published cluster-scale lens that lies in the Vera Rubin Observatory's Legacy Survey of Space and Time Wide-Fast-Deep ``main survey" footprint \citep{lsstcadence}, and as such will be observed frequently over the course of that program. These time delays can be used to further constrain the lens model, and have the potential to be used in constraining cosmological parameters such as the Hubble constant \citep{refsdalH0, napierinprep}.

\subsection{Host Galaxy}
\label{host}

Lensed quasars in principal provide the best view of quasar host galaxies currently possible at redshifts greater than 1 \citep[e.g.,][]{oguri13,sharon17, bayliss17}.  This is because the emission from the very small region of the quasar's engine, which usually dominates the observed flux at most wavelengths, remains a point source when lensed, while the host galaxy is typically distorted into spatially extended arc-like images. Moreover, wide-separation lensed quasars are a particularly attractive target for such studies, as the magnified host galaxy images are unlikely to be conflated with light from the foreground lens (unlike in galaxy-scale lensed quasars), and existing wide-separation systems tend to have higher magnifications than galaxy-scale systems \citep{oguri13,sharon17,2020MNRAS.494.3491L}.

\begin{figure}
     \centering
     \includegraphics[width=241px]{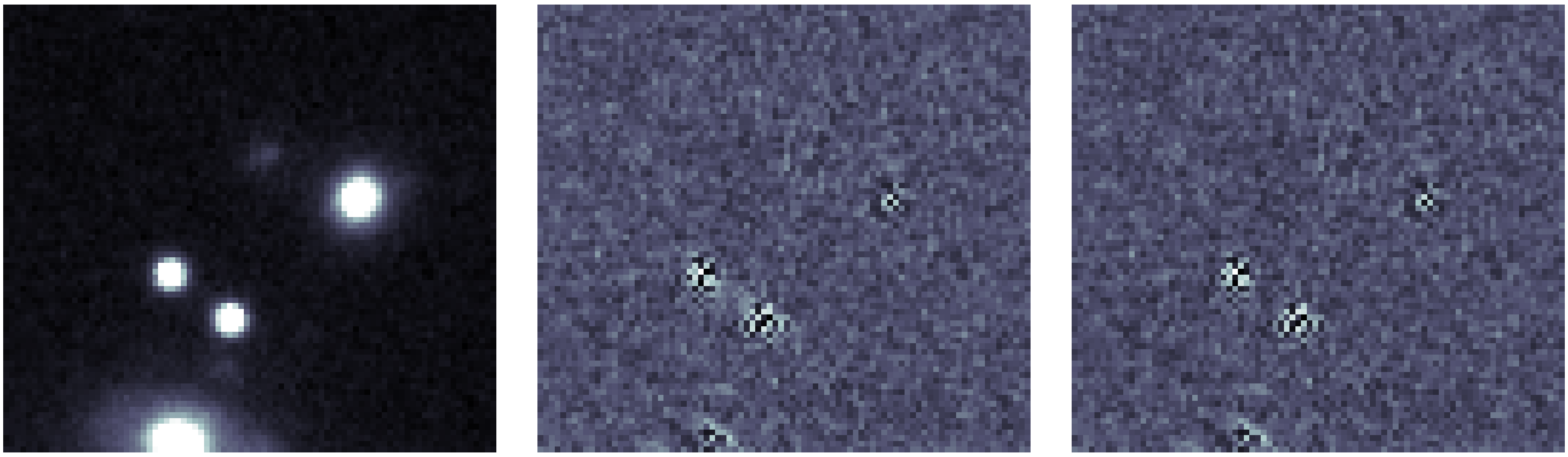}
     \caption{Left to right: 1) LDSS3 $z$-band image of quasar images A and B. 2) Residual from a \texttt{GALFIT} model of this field, using only PSF components at the quasar images. 3) Residual image from a \texttt{GALFIT} model which includes a S\'ersic component located between images A and B. }
     \label{fig:host}
 \end{figure}

We have attempted to extract a detection of the host galaxy in this system. The $grz$ images were each fit with a \texttt{GALFIT} \citep{galfit} model in the region of the two brightest images (A\&B), with a point spread function computed from nearby point sources of similar color, and foreground lens galaxies and other point sources included in the model as needed and indicated by the modeling residuals.
We first fit each quasar image location with a single point source and no component for host galaxy light. In this model, the $z$-band residual indicated possibly significant flux between the two quasar images. No residual flux was seen in the bluer bands. To attempt to measure the host galaxy light in the $z$-band, we added to the model a single S\'ersic component constrained to sit near the midpoint between the two quasar images; the likely image configuration for an extended source is two partial and merging images, and so we expect host flux to be most prominent between the two quasar images.
The $z$-band image region used and the two residual images from this process (one without and one with the additional S\'ersic component) are shown in Figure \ref{fig:host}.

To measure an uncertainty for the notionally fitted host galaxy light, and to establish whether the detection is significant, we added the fitted \texttt{GALFIT} model image to realizations of the noise, and refit the \texttt{GALFIT} model to the resulting image. The noise image was estimated from the RMS of the initial model residual smoothed by a kernel of approximately the same size as the image PSF, with each instance of the noise consisting of that image multiplied by a unitary random Gaussian field. We computed 1000 instances of these model realizations for two cases. 
In the first case we included the notional fitted host galaxy light in the input model image, and in the second we excluded it. In both cases that S\'ersic component was included in the fitted model. If the notional detection is significant, we would expect that the resulting magnitude distribution of the host galaxy component for the first case (notional host light included) would tend brighter than the second case. 
In a set of random comparisons between instances of the two cases above, this occurred 93.8\% of the time, implying that our host galaxy detection is significant at approximately the $2\sigma$ level.
Given these analyses, we measure a $z$-band magnitude of $23.17_{-0.27}^{+0.41}$ for the COOL J0542$-$2125 host galaxy.
Deeper and sharper imaging will be required to refine our understanding of the host galaxy further.

\subsection{Cluster Merger} \label{merger}
The two cluster cores in the direct vicinity of \QSO\, are likely merging \citep[e.g.,][]{turnaround} - assuming both clusters are moving only with the Hubble flow implies a radial separation of $10\pm5$ Mpc (with the range driven by the redshift uncertainties of the two cores) and the tangential separation is much smaller at only 1 Mpc. 
Furthermore, publicly available data from the ROSAT \citep{rosat} satellite suggest X-ray emission in the West cluster's vicinity, but offset from  the cluster center and BCG toward the East cluster.
Such an offset is often present in cluster mergers, such as ``El Gordo" \citep{elgordo}, the ``Bullet Cluster" \citep{bulletdiscovery}, and the ``Baby Bullet" \citep{babybullet}, and has been used to constrain dark matter and alternative cosmologies \citep{bulletDM}. Further X-ray observation would be required to confirm the East and West clusters as one of these mergers.

Another avenue of use for the structure surrounding \QSO\, is presented by the ``Moving Lens Effect" \citep{BG83}, where a transversely-moving potential well induces a small frequency shift in passing photons, from which the transverse velocity can be recovered. Though originally introduced as a higher-order perturbation in the CMB similar to the Rees-Sciama and kinetic Sunyaev-Zeldovich effects \citep{cmb_movinglens}, the effect has been theorized to be observable in closer targets which are lensed by high-transverse-velocity objects. \cite{bullet_velocity} demonstrated that frequency shifts on the order of 1 kms\^{-1} in lensed images behind the Bullet Cluster could be observed, and that lensed quasars are the optimal systems for such a measurement.

The initial measurements presented above will be significantly enhanced by scheduled \textit{Chandra} X-ray and \textit{Hubble} observations (Program ID: 24800144; PI: Napier), and ongoing efforts to measure the time delays of all three lensed quasar images.



\acknowledgments

We would like to express gratitude towards the staff and workers at the 6.5m Magellan Telescopes at the Las Campanas Observatory, Chile for their valuable labor, and in particular for their care and efforts during a global pandemic, during which all of the data in this paper were gathered.

This paper is based on data gathered with the 6.5 m Magellan Telescopes located at Las Campanas Observatory, Chile. Magellan observing time for this program was granted by the time allocation committees of the University of Chicago and the University of Michigan.

This paper is partially based on observations made with the Nordic Optical Telescope, owned in collaboration by the University of Turku and Aarhus University, and operated jointly by Aarhus University, the University of Turku and the University of Oslo, representing Denmark, Finland and Norway, the University of Iceland and Stockholm University at the Observatorio del Roque de los Muchachos, La Palma, Spain, of the Instituto de Astrofisica de Canarias.
The data presented here were obtained in part with ALFOSC, which is provided by the Instituto de Astrofisica de Andalucia (IAA) under a joint agreement with the University of Copenhagen and NOT.

This work is supported by The College Undergraduate program at the University of Chicago, and the Department of Astronomy and Astrophysics at the University of Chicago. 

The Legacy Surveys consist of three individual and complementary projects: the Dark Energy Camera Legacy Survey (DECaLS; Proposal ID 2014B-0404; PIs: David Schlegel and Arjun Dey), the Beijing-Arizona Sky Survey (BASS; NOAO Prop. ID 2015A-0801; PIs: Zhou Xu and Xiaohui Fan), and the Mayall z-band Legacy Survey (MzLS; Prop. ID 2016A-0453; PI: Arjun Dey). DECaLS, BASS and MzLS together include data obtained, respectively, at the Blanco telescope, Cerro Tololo Inter-American Observatory, NSF’s NOIRLab; the Bok telescope, Steward Observatory, University of Arizona; and the Mayall telescope, Kitt Peak National Observatory, NOIRLab. The Legacy Surveys project is honored to be permitted to conduct astronomical research on Iolkam Du’ag (Kitt Peak), a mountain with particular significance to the Tohono O’odham Nation.

NOIRLab is operated by the Association of Universities for Research in Astronomy (AURA) under a cooperative agreement with the National Science Foundation.

This project used data obtained with the Dark Energy Camera (DECam), which was constructed by the Dark Energy Survey (DES) collaboration. Funding for the DES Projects has been provided by the U.S. Department of Energy, the U.S. National Science Foundation, the Ministry of Science and Education of Spain, the Science and Technology Facilities Council of the United Kingdom, the Higher Education Funding Council for England, the National Center for Supercomputing Applications at the University of Illinois at Urbana-Champaign, the Kavli Institute of Cosmological Physics at the University of Chicago, Center for Cosmology and Astro-Particle Physics at the Ohio State University, the Mitchell Institute for Fundamental Physics and Astronomy at Texas A\&M University, Financiadora de Estudos e Projetos, Fundacao Carlos Chagas Filho de Amparo, Financiadora de Estudos e Projetos, Fundacao Carlos Chagas Filho de Amparo a Pesquisa do Estado do Rio de Janeiro, Conselho Nacional de Desenvolvimento Cientifico e Tecnologico and the Ministerio da Ciencia, Tecnologia e Inovacao, the Deutsche Forschungsgemeinschaft and the Collaborating Institutions in the Dark Energy Survey. The Collaborating Institutions are Argonne National Laboratory, the University of California at Santa Cruz, the University of Cambridge, Centro de Investigaciones Energeticas, Medioambientales y Tecnologicas-Madrid, the University of Chicago, University College London, the DES-Brazil Consortium, the University of Edinburgh, the Eidgenossische Technische Hochschule (ETH) Zurich, Fermi National Accelerator Laboratory, the University of Illinois at Urbana-Champaign, the Institut de Ciencies de l’Espai (IEEC/CSIC), the Institut de Fisica d’Altes Energies, Lawrence Berkeley National Laboratory, the Ludwig Maximilians Universitat Munchen and the associated Excellence Cluster Universe, the University of Michigan, NSF’s NOIRLab, the University of Nottingham, the Ohio State University, the University of Pennsylvania, the University of Portsmouth, SLAC National Accelerator Laboratory, Stanford University, the University of Sussex, and Texas A\&M University.

The Legacy Surveys imaging of the DESI footprint is supported by the Director, Office of Science, Office of High Energy Physics of the U.S. Department of Energy under Contract No. DE-AC02- 05CH1123, by the National Energy Research Scientific Computing Center, a DOE Office of Science User Facility under the same contract; and by the U.S. National Science Foundation, Division of Astronomical Sciences under Contract No. AST-0950945 to NOIRLab.

The Pan-STARRS1 Surveys (PS1) and the PS1 public science archive have been made possible through contributions by the Institute for Astronomy, the University of Hawaii, the Pan-STARRS Project Office, the Max-Planck Society and its participating institutes, the Max Planck Institute for Astronomy, Heidelberg and the Max Planck Institute for Extraterrestrial Physics, Garching, The Johns Hopkins University, Durham University, the University of Edinburgh, the Queen's University Belfast, the Harvard-Smithsonian Center for Astrophysics, the Las Cumbres Observatory Global Telescope Network Incorporated, the National Central University of Taiwan, the Space Telescope Science Institute, the National Aeronautics and Space Administration under Grant No. NNX08AR22G issued through the Planetary Science Division of the NASA Science Mission Directorate, the National Science Foundation Grant No. AST-1238877, the University of Maryland, Eotvos Lorand University (ELTE), the Los Alamos National Laboratory, and the Gordon and Betty Moore Foundation.

This research is based on observations made with the Galaxy Evolution Explorer, obtained from the MAST data archive at the Space Telescope Science Institute, which is operated by the Association of Universities for Research in Astronomy, Inc., under NASA contract NAS 5–26555.
%

\vspace{5mm}

\facilities{CTIO/4m Blanco Telescope, Magellan Telescopes 6.5m (Clay/LDSS3C), NOT (ALFOSC), Pan-STARRS1 Telescope}


\software{\texttt{NumPy}
          \citep{numpy_paper}, \texttt{matplotlib} \citep{matplotlib_paper},
          \texttt{astropy} \citep{2013A&A...558A..33A},  
          \texttt{Prospector} \citep{prospector_sps}, 
          \texttt{SourceExtractor} \citep{1996A&AS..117..393B},
          \texttt{Lenstool} \citep{2007NJPh....9..447J}, 
          \texttt{GALFIT} \citep{galfit},
          \texttt{Aperture Photometry Tool}
          \citep{APT}
          }




\bibliography{main}{}
\bibliographystyle{aasjournal}



\end{document}